\documentclass[a4paper, english]{article}
\usepackage{lmodern}
\usepackage[T1]{fontenc}
\usepackage[latin9]{inputenc}
\usepackage{babel}
\usepackage{amsmath,amssymb}
\usepackage{amscd}
\usepackage{amsfonts}
\usepackage{textcomp}
\usepackage{slashed}
\usepackage{cancel}
\usepackage{pb-diagram}
\usepackage{color}
\usepackage[usenames,dvipsnames,svgnames,table]{xcolor}
\usepackage[bookmarks=false,colorlinks=true,linkcolor={red},citecolor=blue,pdfstartview={XYZ null null 1.22},backref,pagebackref]{hyperref}
\usepackage{cleveref}
\renewcommand{\cal}[1]{\mathcal{#1}}
\usepackage{graphicx}
\newcommand{\be}{\begin{equation}}
\newcommand{\ee}{\end{equation}}
\newcommand{\bea}{\begin{eqnarray}}
\newcommand{\eea}{\end{eqnarray}}
\newcommand{\beas}{\begin{eqnarray*}}
\newcommand{\eeas}{\end{eqnarray*}}

\usepackage{soul}
\usepackage{babel}

\def\({\left(}  
\def\){\right)}

\title{Vertical extension of Noether Theorem for Scaling Symmetries}
\begin{document}

\author{J. Antonio Garc\'ia$^1$,  D. Guti\'errez-Ruiz$^2$, \\ R. Abraham Sánchez-Isidro$^3$ \\
\small Departamento de F\'isica de Altas Energ\'ias, Instituto de Ciencias Nucleares\\
\small Universidad Nacional Aut\'onoma de M\'exico,\\
\small Apartado Postal 70-543, Ciudad de M\'exico, 04510, M\'exico\\
\small $^1$ garcia@nucleares.unam.mx,  $^2$daniel.gutierrez@correo.nucleares.unam.mx, \\ \small $^3$abraham.sanchez@correo.nucleares.unam.mx }

\maketitle

\begin{abstract}

The aim of this paper is to present a new approach to construct constants of motion associated with scaling symmetries of dynamical systems. Scaling maps could be symmetries of the equations of motion but not of its associated Lagrangian action. We have constructed a Noether inspired theorem in a vertical extended space that can be used to obtain constants of motion for these symmetries. Noether theorem can be obtained as a particular case of our construction.  To illustrate how the procedure works, we present two interesting examples, a) the Schwarzian Mechanics based on Schwarzian derivative operator and b) the Korteweg-de Vries (KdV)  non linear partial differential equation in the context of the asymptotic dynamics of General Relativity on AdS$_3$. We also study the inverse of Noether theorem for scaling symmetries and show how we can construct and identify the generator of the scaling transformation, and how it works for the vertical extended constant of motion that we are able to construct. We find an interesting contribution to the symmetry associated with the fact that the scaling symmetry is not a  Noether symmetry of the action. Finally, we have contrasted  our results with recent analysis and previous attempts to find constants of motion associated with these beautiful scaling laws. 

\end{abstract}

\section{Introduction}

Given the importance of scaling symmetries in many areas of physics like hydrodynamics, classical mechanics \cite{Landau}, renormalization, thermodynamics, phase transitions, critical phenomena, and non relativistic conformal field theories  \cite{Henriksen,Sedov}, an interesting question is if we can apply Noether theorem to these type of symmetries and obtain an associated conserved quantity. This question has been addressed recently \cite{Horvathy} (to see previous attempts the reader is referred to \cite{Takahisa}), and a modified Noether theorem has been constructed. The problem to apply the Noether theorem to scaling symmetries is that these kind of symmetries are NOT Noether symmetries, they change the Lagrangian by a numerical factor times the Lagrangian itself, and a total derivative with respect to time. We need to modify Noether theorem to allow these type of symmetries to enter in the framework of the Noether theorem. 
Nevertheless, the scaling infinitesimal symmetries are {\em symmetries of the equation of motion} in the sense that they satisfy the equation of variations\footnote{Another approach is to work with these type of symmetries in the context of non-Noetherian symmetries (also called s-equivalent symmetries) but we will not follow this approach here. For details, we will refer the reader to \cite{Hojman}.}.

Unfortunately, the price to pay in the construction presented in \cite{Horvathy} is quite high. This ``generalized Noether theorem'' applied to scaling symmetries has implicit that we know the complete solution of the system to construct an associated constant of motion. At first sight, all the attempt appears as a practical non sense, but nevertheless the construction can be used to learn interesting properties of systems with scaling symmetries. In fact, the construction is closely related with Hamilton-Jacobi theory from classical mechanics. One open question is if the modified Noether theorem for scaling infinitesimal symmetries admits an inverse like Noether theorem. To be precise, if the conserved quantity associated with the scaling infinitesimal symmetry can be used to reconstruct the scaling infinitesimal symmetry. In fact, we can identify the generator of the scaling symmetry, but as the generator is NOT a constant of motion of the system, the resulting symmetry is non-Noether. In this paper, we will show that the inverse of Noether theorem applied to the conserved quantity obtained in \cite{Horvathy} produces another symmetry,  {\em different} from the scaling symmetry. This is another way to see that the scaling symmetry is non-Noether.  

Next, we will present a central result of this paper, a complete new perspective to find conserved quantities associated with non-Noether scaling symmetries. Our procedure consists of two steps. First, we will use the formalism of the vertical extension of a Lagrangian system \cite{Sardanashvily}. This extension has the advantage of working even when the symmetries are non Noether, i.e. symmetries of the equations of motion that are not symmetries of the action. In fact, the aim of the vertical extension is to have a conceptual framework to include in the same footing 
the Lagangian variational equations of motion and the Jacobi equations. The symmetries of the system are controlled by these fundamental Jacobi equations, in the sense that the solutions of these equations, also know as the equations of variations, are symmetries of the equations of motion. Some of these symmetries could also be Noether symmetries, so they are also symmetries of the Lagrangian action. Second, we will use this vertical extension for the specific case of scaling symmetries that are in general symmetries of the equations of motion but NOT symmetries of the action to construct a Noether theorem. Fortunately, we can construct a Noether inspired theorem that 
allows to find a conserved quantity associated with non Noetherian scaling symmetries. Our procedure does NOT need that we know the complete solution of the dynamical system {\em a priori} to construct the associated conserved quantity! In this way, we have a crucial improvement over the original proposal presented in \cite{Horvathy}. 

We will present two interesting non trivial examples in the context of the Noether theorem for scaling symmetries and from the point of view of the vertical extension. The first example is a dynamical system constructed from a powerful $SL(2,R)$  symmetry based on the Schwarzian derivative operator. 
Because $sl(2, R)$ is a finite-dimensional subalgebra of the Virasoro algebra, the Schwarzian derivative arises naturally within the context of string theory. In recent years, there has been much activity in studying 1d quantum mechanics that arises as the low energy limit of the solvable theory with maximally chaotic behaviours $-$ the so called Sachdev-Ye-Kitaev model. A peculiar feature of the system is that its Lagrangian density is proportional to the Schwarzian derivative of a specific function.

Schwarzian mechanics is a classical mechanical model that in some sense mimics some of the properties of this interesting dynamical system. It is a system
which is governed by the third order equation of motion,
${\rm Schw}(\rho(t))=\lambda,$ 
where $\lambda$ is a constant and ${\rm Schw}$  is the Schwarzian derivative operator .
The Hamiltonian formulation is unconventional and a Lagrangian formulation was missing until the work presented in \cite{Galajinsky}. We found that this Lagrangian formulation has an interesting anisotropic scaling that is not a symmetry of the action but is a symmetry of the equation of motion. A very interesting relation between this model and the de Alfaro-Fubini-Furlan (AFF) conformal mechanics \cite{Alfaro} was found recently in \cite{Filyukov}. We believe that this result could be used to map constants of motion of the Schwarzian mechanics associated with scaling symmetry to constants of motion of the AFF conformal mechanics \cite{wp}.

The second example comes from the work \cite{Grumiller} on asymptotic symmetries of the bulk gravity in $AdS_3$ with modified boundary conditions that generalizes the boundary conditions of Brown-Henneaux \cite{BH}  to associate asymptotic symmetries of the bulk gravity action in $AdS_3$ with symmetries of a CFT in $1+1$. 
The modified boundary condition gives rise to KdV equations. Notice that we have instead an integrable nonrelativistic  equation with
an infinite set of commuting conserved charges, implying the integrability of the system.


This beautiful discovery was also made for the so called potential modified KdV partial differential equations (pmKdV), but in this case, the asymptotic symmetries correspond to anisotropic Lifshitz scalings with dynamical exponent $z$ that are Noether symmetries. So, for our proposes the interesting case is the non-linear partial differential evolution equation corresponding to the well known KdV dynamical system, because in that case we have anisotropic scaling that is a symmetry of the equation of motion but not of the Lagrangian action. A unified approach of the two cases, pmKdV and KdV, is presented in \cite{Ojeda}.

In ref. \cite{Grumiller}, the authors suggest that the scaling properties of the KdV Lagrangian action can be worked out in terms of the generalized Noether theorem for scaling symmetries \cite{Horvathy}. We will show here that the correct point of view to attack this problem is by using our theory developed in section 3. In fact, we were unable to work this interesting example in the context of \cite{Horvathy}.  

We will present the two examples in the context of generalized Noether theorem for scaling symmetries and in the context of the vertical extension, so that the reader can contrast the advantages or inconveniences of one approach or the other.

Scaling symmetries are also quite relevant in the construction of the recent studies on non relativistic AdS/CFT, Horava gravity, anisotropic mechanics, Scrhoedinger group, Bargmann and Carroll groups, and Newton-Cartan algebras.

\section{Review of the generalized Noether theorem for scaling symmetries}

In this section, we will review the central idea of the generalized Noether theorem for scaling symmetries, and present a new result using the inverse of Noether theorem to construct a symmetry associated with the corresponding constant of motion. As the scaling symmetry is a non Noether symmetry, it is not generated by a constant of motion. The question is what is the symmetry associated with the constant of motion obtained via this generalization of Noether theorem. It is clear that it is NOT the scaling symmetry.

Consider a scaling mapping given by
$$q^i\to \lambda^a q^i, \qquad t\to \lambda^b t,$$
where $a,b$ are some given constants. The corresponding infinitesimal transformation
\begin{equation}\label{delta-q}
\Delta_s q^i=a q^i - b \dot q^i t
\end{equation}
is a symmetry of the  equations of motion if the forces satisfy (see appendix)
\begin{equation}\label{force-cond}
-bt\frac{\partial F^i}{\partial t}+(a-2b)F^i-(a-b)  \frac{\partial F^i}{\partial\dot q^j}\dot q^j-\frac{\partial F^i}{\partial q^j} a q^j=0.
\end{equation}
The infinitesimal transformation (\ref{delta-q}) is a generalized Lagrangian symmetry if we can find a number $\Lambda$ and a function $f$ such that
\be\label{def-sca}
\Delta L=\Lambda L(q,\dot q ,t)+\frac{d f}{dt}.
\ee
If the number $\Lambda$ is zero, the scaling mapping (\ref{delta-q}) is a Noether symmetry associated to the Lagrangian $L$. The number $\Lambda$ depends on the Lagrangian description of the dynamical system and is a function of the constants $a,b$, $\Lambda=\Lambda(a,b)$.

Now, according to \cite{Horvathy}, it is still possible to find a constant of motion associated with the generalized symmetry $\Lambda\not=0$ given by
\be\label{ch}
C_S=\frac{\partial L}{\partial \dot q^j} (a q^j-b\dot q^j t)-f-\Lambda S(t),
\ee
where $S(t)$ is the Lagrangian action
$$S=\int L dt.$$
Of course, if we want to know explicitly the action functional (and then the constant of motion $C_S$), we need to know the complete solution of the system according to the boundary data of the variational principle. Nevertheless, it is interesting to observe that the on shell derivative, hereafter denoted with a bar, of $C_S$ is in fact zero\footnote{Our notation is $$
\frac{{\bar d} C}{dt}=F^i\frac{\partial C}{\partial\dot q^i}+ \dot q^i\frac{\partial C}{\partial q^i}+\frac{\partial C}{\partial t}.$$}
$$\frac{\bar d C_S}{dt}=0,$$
by using the Lagrangian version of the Hamilton-Jacobi equation \cite{Landau}
$$\frac{\bar d S}{dt}=L,$$
which implies
$$\frac{\bar d }{dt}(\frac{\partial L}{\partial \dot q^j} (a q^j-b\dot q^j t)-f)=\Lambda L.$$
The apparent non locality in the definition of the constant $C_S$ disappears when we consider $S$ in (\ref{ch}) as a function of $q,\dot q,t$ and NOT as a functional of the dynamical trajectory of the system.

Here we found an apparent paradox: if we use the constant of motion $C_S$ to calculate a symmetry $\Delta q^i$ using the inverse of Noether theorem, we found a new symmetry 
\begin{equation}\label{Inv-NT}
\Delta_S q^i =W^{ij}\frac{\partial C_S}{\partial\dot q^j},
\end{equation}
where $W^{ij}$ denotes the inverse of the Lagrangian Hessian matrix $W_{ij}=\frac{\partial^2L}{\partial\dot{q}^i\partial\dot{q}^j}$ (see Appendix A2). This symmetry is a Noether symmetry (by construction) and does NOT coincide with the scaling symmetry (\ref{delta-q}). In fact, we found that the difference between the new symmetry and the scaling symmetry is
$$\Delta_S q^i-\Delta_s q^i=-\Lambda W^{ij}\frac{\partial S}{\partial \dot q^j}.$$
This is one of the results of this paper. In the following lines we will present a proof of our result. To this end, define
$$G=\frac{\partial L}{\partial \dot q^j} (a q^j-b\dot q^j t)-f.$$
With this definition, the constant of motion $C_S$ can be written as
\be\label{def-ch}
C_S=G-\Lambda S.
\ee
Now we can show that $G$ is the generator of the scaling symmetry using the inverse of the Noether theorem (\ref{Inv-NT})
$$\Delta_s q^i=W^{ij}\frac{\partial G}{\partial \dot q^j}. $$
For that end define
\begin{equation}\label{Gs}
G=\frac{\partial L}{\partial \dot q^j} (a q^j-b\dot q^j t)-f
\end{equation}
With this definition the constant of motion $C_S$ can be written as
\be\label{def-ch}
C_S=G-\Lambda S
\ee
Now we will show that $G$ is the generator of the scaling symmetry using the inverse of the Noether theorem (\ref{Inv-NT})
$$\Delta_s q^i=W^{ij}\frac{\partial G}{\partial \dot q^j} $$
where $W_{ij}$ is defined as
$$W_{ij}=\frac{\partial^2 L}{\partial\dot q^i\partial\dot q^j}$$
and $W^{ij}$ is the inverse of $W_{ij}$.
For that end we will take the scaling transformation (\ref{delta-q}) and find the general form of its generator
$$W_{ij}(a q^i-b\dot q^i t) =\frac{\partial G_s}{\partial \dot q^j}$$
for some $G_s$. As a consequence, $G_s$ must be of the form
$$G_s=\frac{\partial L}{\partial \dot q^i} (a q^i-b\dot q^i t)-f,$$
for some function $f$ to be determined below.  It is straightforward to see that
 $$-b t \frac{\partial L}{\partial \dot q^i}-  \frac{\partial f}{\partial \dot q^i}=0$$
 must be fulfilled, which in turn implies 
$$f=-L\delta t.$$
But this is the  function $f$ that we need in order for the scaling transformation to be a symmetry as defined in eq (\ref{def-sca}). Needless to say that the generator $G_s$ coincide with the generator $G$ defined in (\ref{Gs}).

As the generator $G$ of the scaling symmetry is not a constant o motion the symmetry is not a Noether symmetry.


Summarizing, using the inverse Noether theorem for the constant of motion $C_S$
and asking for the form of the corresponding Noether symmetry associated with this constant of motion, we found a new symmetry  that differs from the scaling symmetry by the term
$-\Lambda W^{ij}\frac{\partial S}{\partial \dot q^j}$.
Notice that here, $S$ is a function of $q,\dot q, t$.
We do not advance any physical interpretation of this new symmetry, but the constant of motion $C_S$ can be related with initial data by
$$ C_S=aq^i(0)\dot q^i(0).$$
This constant of motion has an interesting physical interpretation. For example, in the case of the Kepler problem that was considered in \cite{Horvathy, Horvathy-1}, $C_S$ is related with the conservation of the Liouville flow in phase space. $C_S$ is also related with two solutions of the Hamilton-Jacobi equation that differ by a Legendre transformation and to the virial theorem in classical mechanics.

The other remarkable result is already presented in eq (\ref{def-ch}). We found that the Lagrangian of a scale invariant theory is always the on shell derivative with respect to time of the generator of the scaling symmetry
\be\label{G-eq}
\frac{\bar d G}{dt}=\Lambda L.
\ee
This last equation can be seen as a consequence of the scaling symmetry $\Delta_s q^i$  of the Lagrangian $L(q,\dot q,t)$. It defines the scaling symmetry. But on the other hand, we now have from the definition of the action $S$ that
\be\label{S-eq}
\frac{\bar d S}{dt}=L.
\ee
This equation does not the depend on the scaling symmetry.
So, given $G$ as the generator of the scaling symmetry, the equations (\ref{G-eq}) and (\ref{S-eq}) imply that its difference $G-\Lambda S$ is a conserved quantity, i.e. a constant of motion of the system,
$$ G-\Lambda S\thicksim C$$
and this is the content of the Noether theorem for scaling symmetries.

The problem for the explicit construction of the constant of motion $C_S$ is that we need the complete solution of the dynamical system.  Some hints on the form of the function $C_S$ could be guessed from its well defined scaling properties ($C_S$ has dimensions of action and therefore scales as the action scales) or/and the corresponding Hamilton-Jacobi theory. From $C_S$ we can reconstruct the Lagrangian action $S$ as the generator of the canonical transformation that changes the variables $q^i,p_i$ to $Q^i,P_i$ where $Q's$ and $P's$ are constants of motion. It could be of interest to find more properties associated with the new symmetry presented here and the extensions of these ideas to field theory.

\section{Vertical extension and the generalized Noether theorem for scaling maps}

We will propose a new approach to relate a constant of motion to scaling symmetries. First, we notice that an interesting setup to study the Noether theorem for scaling maps is the formalism of vertical extension of Lagrangian dynamics. This formalism has been developed in  \cite{Sardanashvily}, and has the advantage to include in the same footing the dynamics of the degrees of freedom $q^i$ and the symmetries of the equations of motion represented by the Jacobi fields (for details see \cite{Sardanashvily}).

The construction of this formalism for Lagrangian mechanics starts from a well defined regular Lagrangian $L(q,\dot q, t)$ in tangent space with coordinates $(q^i,\dot q^i)$, $i=1,\ldots, n$, and its vertical extension 
$(q^i,\dot q^i,\eta^i,\dot\eta^i),$ 
where $\eta^i$ and $\dot\eta^i$ are the so called Jacobi Fields associated to the Jacobi equation of variations (see below). Any solution of the Jacobi equations is a symmetry of the equations of motion associated with the given Lagrangian $L(q,\dot q,t)$.
The vertical extension of the Lagrangian $L$ is defined by\footnote{This definition of $\gamma$ has the following property. First identify $\eta$ with $\delta q$. We will call it a projection from vertical extended space to configuration space. The projection is then $\gamma=\delta L$. In terms of the projected  $\gamma$ Noether theorem is just 
$EL(\gamma\Big |_{\eta=\delta q})\equiv 0$. If the symmetry is non-Noether $\gamma\Big |_{\eta=\delta q}$ is an s-equivalent Lagrangian (for details see \cite{Hojman}).}

\begin{equation}\label{gamma}
\gamma(q,\eta,\dot q,\dot\eta)=\frac{\partial L}{\partial\dot q^i}\dot\eta^i+\frac{\partial L}{\partial q^i}\eta^i.
\end{equation}
Notice that
\begin{equation}
\frac{\partial \gamma}{\partial\dot \eta^i}= \frac{\partial L}{\partial\dot q^i}, \quad \frac{\partial \gamma}{\partial \eta^i}= \frac{\partial L}{\partial q^i},
\end{equation}
so $\gamma$ can be written as
\begin{equation}
\gamma(q,\eta,\dot q,\dot\eta)=\frac{\partial \gamma}{\partial\dot \eta^i}\dot\eta^i+\frac{\partial \gamma}{\partial \eta^i}\eta^i.
\end{equation}
Another equivalent expression for  $\gamma$ is
\begin{equation}\label{gamma-eq-mot}
\gamma=-\eta^i ( \text{EM}^q_i(L))+\frac{d}{dt}\left(\frac{\partial L}{\partial\dot q^i}\eta^i\right),
\end{equation}
where $ \text{EM}^q_i(L)$ are the Euler-Lagrange equations of motion associated with the given Lagrangian $L$
\begin{equation}
 \text{EM}^q_i(L)=\frac{d}{dt}\frac{\partial L}{\partial\dot q^i}-\frac{\partial L}{\partial q^i}.
\end{equation}
Now, a general infinitesimal variation of $\gamma$ is
\begin{equation}\label{delta-gamma}
\Delta\gamma=-\Delta q^i ( \text{EM}^q_i(\gamma))-\Delta\eta^i ( \text{EM}^\eta_i(\gamma)) +  \frac{d}{dt}\left(\frac{\partial \gamma}{\partial\dot q^i}\Delta q^i  + \frac{\partial \gamma}{\partial\dot \eta^i}\Delta \eta^i\right),
\end{equation}
where
\begin{equation}
 \text{EM}^\eta_i(\gamma)=\frac{d}{dt}\frac{\partial \gamma}{\partial\dot \eta^i}-\frac{\partial \gamma}{\partial \eta^i}
\end{equation}
are the equations of motion associated to the Lagrangian $L$, and $ \text{EM}^q_i(\gamma)$ are the Jacobi equations that can be explicitly written as
\begin{equation}
W_{ij}(q,\dot q,t)\ddot\eta^j+N_{ij}(q,\dot q,t)\dot\eta^j+M_{ij}(q,\dot q,t)\eta^j=0,
\end{equation}
with the notation given in the appendix B. Notice that a solution of the Jacobi equation is for the Jacobi fields $\eta^i$ as functions of $q^i,\dot q^i,t$ and not for the variations $\Delta q, \Delta\eta$. This point is quite relevant to obtain projections from the extended space to the configuration space. The variations  $\Delta q, \Delta\eta$ can be defined by its action on $\gamma$ as follows.

Consider a scaling mapping in the configuration space 
\begin{equation}\label{scaling-sym}
\Delta_s q^i=a q^i-b\dot q^i t,
\end{equation}
where the parameters $a,b$ are some constants associated with the anisotropic scaling
$$
q^i\to \lambda^a q^i,\qquad t\to\lambda^b t.
$$
 As we already noticed the variation of $L$ along this symmetry is given by 
\begin{equation}
\Delta_s L=\Lambda(a,b)L-\frac{d}{dt} (L\delta t).
\end{equation}
Then we propose an extension of the scaling symmetry (\ref{scaling-sym}) to the vertical space given by
\begin{equation}\label{ext-sym}
\Delta_s q^i=a q^i-b\dot q^i t, \quad \Delta_s \eta^i=a \eta^i-b\dot \eta^i t.
\end{equation}
The action of this symmetry over $\gamma$ is
\begin{equation}\label{delta-gamma_s}
\Delta_s \gamma=\Lambda(a,b)\gamma-\frac{d}{dt} (\gamma\delta t),
\end{equation}
with $\Lambda(a,b)$ some function of the scaling parameters $a,b$. The action of the extended scaling symmetry on $\gamma$  defines the extension of the symmetry to the vertical space.  

We will call the extended symmetry (\ref{ext-sym}) and its action over $\gamma$ (\ref{gamma}) the vertical extension of the scaling symmetry $\Delta_s q^i$ (\ref{scaling-sym}). Our definition of the extended symmetry and its action over $\gamma$ is inspired in the same arguments as the construction of the scaling mapping and the scaling properties of the Lagrangian $L$. It is easy to show that the scaling properties of $\gamma$ are the same as the scaling properties of the original Lagrangian $L$.

Now, specializing $\Delta \gamma$ given in (\ref{delta-gamma}) to the scaling symmetry (\ref{ext-sym}), we can construct $\Delta_s\gamma$
\begin{equation}\label{delta-scaling-gamma-1}
\Delta_s\gamma=-\Delta_s q^i ( \text{EM}^q_i(\gamma))-\Delta_s\eta^i ( \text{EM}^\eta_i(\gamma)) +  \frac{d}{dt}\left(\frac{\partial \gamma}{\partial\dot q^i}\Delta_s q^i  + \frac{\partial \gamma}{\partial\dot \eta^i}\Delta_s \eta^i\right).
\end{equation}
Using  (\ref{gamma-eq-mot})
 we can write (\ref{delta-gamma_s}) in the form
\begin{equation}\label{delta-scaling-gamma-2}
\Delta_s \gamma=\Lambda(a,b)
\left(-\eta^i ( \text{EM}^q_i(L))+\frac{d}{dt}\left(\frac{\partial L}{\partial\dot q^i}\eta^i\right)\right)
-\frac{d}{dt} (\gamma\delta t).
\end{equation}
Equating the relation given in (\ref{delta-scaling-gamma-1}) with this last result (\ref{delta-scaling-gamma-2}) we obtain
\begin{equation}\label{fund-rel}
-\Delta_s q^i ( \text{EM}^q_i(\gamma))-\Delta_s\eta^i ( \text{EM}^\eta_i(\gamma)) +  \frac{d}{dt}\left(\frac{\partial \gamma}{\partial\dot q^i}\Delta_s q^i  + \frac{\partial \gamma}{\partial\dot \eta^i}\Delta_s \eta^i\right)=
\end{equation}
$$\Lambda(a,b)\left(- \eta^i ( \text{EM}^q_i(L)+\frac{d}{dt}(\frac{\partial L}{\partial\dot q^i}\eta^i)\right)
-\frac{d}{dt} (\gamma \delta t).$$

From here, we can read a conserved quantity in the vertical extended space 
\begin{equation}\label{const-mot-ext}
 C^E_s = \frac{\partial \gamma}{\partial\dot q^i}\Delta_s q^i  + \frac{\partial \gamma}{\partial\dot \eta^i}\Delta_s \eta^i-\Lambda\frac{\partial L}{\partial\dot q^i}\eta^i+\gamma\delta t.
\end{equation}
Noticing that
\begin{equation}
\frac{\partial\gamma}{\partial \dot q^i}=\frac{\partial L}{\partial\dot q^i\partial\dot q^j}\dot\eta^j+\frac{\partial L}{\partial\dot q^i\partial q^j}\eta^j
\end{equation}
and
\begin{equation}
\frac{\partial \gamma}{\partial\dot \eta^i}= \frac{\partial L}{\partial\dot q^i},
\end{equation}
the constant of motion (\ref{const-mot-ext}) in terms of the original Lagrangian function can be written as
\begin{equation}
 C^E_s = \left(\frac{\partial L}{\partial\dot q^i\partial\dot q^j}\dot\eta^j+\frac{\partial L}{\partial\dot q^i\partial q^j}\eta^j
\right)\Delta_s q^i  -\Lambda\frac{\partial L}{\partial\dot q^i}\eta^i + \frac{\partial L}{\partial\dot q^i}\Delta_s \eta^i+ \gamma\delta t,
\end{equation}
where $\gamma$ is just the original definition (\ref{gamma}).

This is the main result our paper. We want to remark that this result is quite surprising. The constant of motion is based on the scaling symmetry (\ref{ext-sym}) and we not need to use the Lagragian action, as in the generalized Noether theorem disused in our previously (see section 2). It seems to be a very interesting and powerful result. To obtain dynamical information in the $(q,\dot q)$ space from this constant of motion, we need to {\em project} it to configuration space.  It is worth noting that a particular solution of the Jacobi equation is a symmetry in the configuration space, $\eta^i=\eta^i(q,\dot q ,t)$. Remarkably, we can use this symmetry to define the projection needed and we can use {\em any} solution of the Jacobi equation for that end. In this way, we can associate to any symmetry (not just a Noether symmetry) a constant of the motion in configuration space, generalizing the standard Noether construction. Here, we want to remark that our construction of the constant of motion is based on the scaling symmetry given in (\ref{ext-sym}).

Now we have the following theorem: Given a constant of motion in extended space $C^E_s$, and a particular solution of the Jacobi equations $\eta^i$, the restriction of $C^E_s$ to the subspace $q^i,\dot q^i$ is a constant of motion in configuration space
\begin{equation}\label{const-config}
C^E_s\Big|_{\eta^i}=C^q(q,\dot q,t).
\end{equation}

Observe that in the basic relation (\ref{fund-rel}), the equations of motion $ \text{EM}^\eta_i(\gamma)$ are the same as the equations of motion $ \text{EM}^q_i(L)$, so we can write
\begin{equation}\label{der-t-const-mot-ext}
\frac{d}{dt}C^E=\frac{d}{dt}\left(\frac{\partial \gamma}{\partial\dot q^i}\Delta_s q^i  + \frac{\partial \gamma}{\partial\dot \eta^i}\Delta_s \eta^i-\Lambda\frac{\partial L}{\partial\dot q^i}\eta^i+\gamma \delta t\right)=
\end{equation}
$$\Delta_s q^i ( \text{EM}^q_i(\gamma))+(\Delta_s\eta^i- \Lambda\eta^i) ( \text{EM}^q_i(L)).
$$


To prove  this theorem, just observe that the derivative with respect to time of $C^E_s$ is\footnote{Here, we are restricting ourselves to the case where the equations of motion in configuration space and the equations of motion of the Jacobi fields are independent. For an example where this condition is not fulfilled, see below.} 
\begin{equation}\label{time-der-const-mot-ext-1}
\frac{d}{dt}C^E_s=\frac{\partial C^E_s}{\partial\dot q^i} \text{EM}_i+\frac{\partial C^E_s}{\partial\dot \eta^i}\text{EMJ}_i.
\end{equation}
Here, we use a simpler notation where $ \text{EM}_i$ are the equations of motion in the {\em configuration space} and $\text{EMJ}_i$ are the Jacobi equations. 
As the time derivative commutes with the projection $\eta=\eta(q,\dot q, t)$ \cite{Henneaux}, we have
\begin{equation}
\frac{d }{dt}(C^E_s)\Big|_{\eta^i} = \frac{d }{dt}\left(C^E_s\Big|_{\eta^i}\right)=\frac{d}{dt}C^q(q,\dot q,t),
\end{equation}
so we can conclude that the derivative with respect to time of the projected constant of motion (\ref{const-config}) is
\be
\frac{d}{dt}C^{q}= (\Delta_s\eta^i-\Lambda\eta^i)\Big|_{\eta^i(q,\dot q,t)} \text{EM}^q_i .
\ee
This proves our theorem. This result is crucial to extract dynamical information to integrate the equations of motion in the configuration space. Another interesting observation is that the projection of the scaling symmetry $\Delta_s\eta^i\Big|_{\eta^i(q,\dot q,t)}$ is not generated by a constant  of motion in the configuration space. Nevertheless, the corrected symmetry
\be
(\Delta_s\eta^i-\Lambda\eta^i)\Big|_{\eta^i(q,\dot q,t)}
\ee
is generated by a constant of motion $C^q$ and is a Noether symmetry in the configuration space. This argument follows from the inverse of Noether theorem.
Interestingly enough, by comparison of the equations (\ref{der-t-const-mot-ext}) and (\ref{time-der-const-mot-ext-1}), we can obtain the inverse of Noether theorem for scaling symmetries in extended space
\begin{equation}
\frac{\partial C^{(q,\eta)}}{\partial\dot \eta^i}=W_{ij}\Delta_s q^i, \quad \frac{\partial C^{(q,\eta)}}{\partial\dot q^i}=W_{ij}(\Delta_s \eta^i-\Lambda \eta^j).
\end{equation}
Surprisingly, the second relation produces a {\em new symmetry}  that does not coincide with the starting symmetry of the Jacobi fields proposed in  (\ref{ext-sym}), but comes corrected by the term $\Lambda \eta$ and produces a well defined symmetry of the equations of motion for the Jacobi fields.

Just to summarize what we have achieved so far, we have constructed a constant of motion in a vertical extended space by doubling the degrees of freedom to include the Jacobi fields $\eta$ along with the configuation space variables $q^i$. The construction of this constant of motion is based on the extended scaling symmetry proposed in (\cite{Horvathy}).
Now, to extract dynamical information in configuration space from this constant of motion in the extended space, we need a projection to configuration space. This projection can be performed using any particular solution of the Jacobi equations. The solution of these Jacobi equations are symmetries of the equations of motion in configuration space, $\eta=\eta(q,\dot q, t)$. This symmetry can be a Noether symmetry or just a symmetry of the equations of motion, but not of the Lagrangian action. In either case, we have shown that our formalism gives a constant of motion based on the scaling symmetry and a particular solution of the Jacobi equation. So, in some sense, our procedure associates a constant of motion to two symmetries, but one of them is given by construction, the scaling symmetry.

To see how this vertical extension works, let us consider the Classical Mechanics of a Lagrangian of the form $L=T-V$ and check if we have a scaling transformation that is a symmetry of the equations of motion but not of the action. 
To that end, consider a Lagrangian $L=\frac12 {\dot q}^2-V(q)$.  The associated vertical extension is
\be
\gamma=\dot q^i\dot\eta^i-\frac{\partial V}{\partial q^i}\eta^i,
\ee
and the conserved quantity is
\be
C^E_s=\dot\eta^i\Delta q^i-\Lambda \dot q^i\eta^i+\dot q^i\Delta\eta^i+\dot q^i\dot \eta^i\delta t -\frac{\partial V}{\partial q^i}\eta^i\delta t,
\ee
which, substituting the scaling symmetries
\be\label{ssym} 
\Delta_s q^i=a q^i-b \dot q^i\delta t,\qquad \Delta_s \eta^i=a \eta^i-b \dot \eta^i\delta t,
\ee
reduces to
\be
C^E_s=a\dot\eta^i q^i-a\dot q^i\eta^i-b\dot\eta^i\dot q^i t +b\dot q^i\eta^i-b t\frac{\partial V}{\partial q^i}\eta^i.
\ee
Now it is easy to show that if $V(\mu q)=\mu^kV(q)$, then $C^E_s$ is conserved and the scaling parameters are $a$ and $b=a(1-k/2)$. This in turn implies that $\Lambda=2a-b=a(1+k/2)$. 
This observation coincides with the result obtained in \cite{Horvathy}. In the particular case $a=2,b=3$, we obtain in the vertical extended space, the analog of the beautiful result associated with the Kepler third law \cite{Horvathy,Horvathy-1}.

The application of the inverse of the Noether theorem to the constant of motion in the extended space $C^E_s$ gives
\be
\frac{\partial C^E_s}{\partial\dot \eta^i}=a q^i-b t\dot q^i,
\ee
which coincides with equation (\ref{ssym}), and
\be
\frac{\partial C^E_s}{\partial\dot q^i}=-a\eta^i-b t\dot\eta^i+b\eta^i,
\ee
corresponds to $\Delta_s \eta^i=a \eta^i-b t \dot \eta^i$ and $\Lambda=2a-b$, as expected from our general analysis.

\section{Examples}

We will apply our methods to two interesting and non trivial examples, the Schwarzian Mechanics \cite{Galajinsky} and the KdV equation \cite{Grumiller}.

\subsection{Schwarzian mechanics}
The action of Schwarzian mechanics is \cite{Galajinsky}
\begin{equation}
S=-\frac{1}{4}\int\mathrm{d}t\,\frac{\ddot{\rho}^{2}}{\dot{\rho}^{2}}.
\end{equation}
The variation of the action is
\be
\Delta S=-\frac14\int\mathrm{d}t
\left\{
\left(2\frac{d^2}{dt^2}\left(\frac{\ddot \rho}{\dot \rho^2}\right)+
 2\frac{d}{dt}\left(\frac{\ddot \rho^2}{\dot \rho^3}\right) 
\right)\delta\rho
+\frac{d G}{dt}
\right\},
\ee
where
\be
G=2\left(\frac{\ddot \rho}{\dot \rho^2}\right)\delta\dot\rho-
2\left(\frac{\ddot \rho^2}{\dot \rho^3}\right)\delta\rho-
2\frac{d}{dt}\left(\frac{\ddot \rho}{\dot \rho^2}\right)\delta\rho.
\ee
The equation of motion is
\be
-\frac{1}{2\dot\rho^2}\ddddot\rho+2\frac{\ddot{\rho}}{\dot\rho^3}\dddot\rho-\frac32\frac{\ddot\rho^3}{\dot\rho^4}=0,
\ee
which can also be written in the form
\begin{equation}
-\frac{1}{2\dot{\rho}}\frac{\mathrm{d}}{\mathrm{d}t}\mathrm{Schw}\rho=0.\label{eqmotion}
\end{equation}
where
\be
{\rm Schw}(\rho(t))=\frac{\dddot{\rho}(t)}{\dot{\rho}(t)}-\frac32\left(\frac{\ddot{\rho}(t)}{\dot{\rho}(t)}\right)^2
\ee
is the Schwarzian derivative.
The infinitesimal Noether symmetries associated with the Schwarzian derivative are 
\be
\rho'(t) = \rho(t) + a, \quad \rho'(t) = \rho(t) + \beta\rho(t), \quad \rho'(t) = \rho(t) + \gamma\rho(t)^2,
\ee
related with the sl(2,R) generators that can be constructed from the Noether theorem, 
\begin{equation}
 \text{EM}\Delta\rho+\frac{\mathrm{d}}{\mathrm{d}t}(G-F)=0,
\end{equation}
where $ \text{EM}$ is the equation of motion and  $\Delta\rho=\delta\rho-\dot{\rho}\delta t$ is the Noether symmetry of the action. Here, $F$ is a function such that 
$\Delta L=\frac{\mathrm{d}F}{\rm{d}t}$, and $G$ is
\begin{equation}
G=\left(\frac{\partial L}{\partial\dot{\rho}}-\frac{\mathrm{d}}{\mathrm{d}t}\frac{\partial L}{\partial\ddot{\rho}}\right)\Delta\rho+\frac{\partial L}{\partial\ddot{\rho}}(\Delta\rho)^{\cdot}.
\end{equation}
So we conclude that $C=G-F$ is a constant of motion associated with the given symmetry.
\begin{enumerate}
\item Invariance under translations: $\Delta\rho=1$, $F=0$, and the conserved
quantity is the momentum
\begin{equation}
C=\frac{\dddot{\rho}}{2\dot{\rho}^{2}}-\frac{\ddot{\rho}^{2}}{2\dot{\rho}^{3}}=\frac{1}{2\dot{\rho}}\left(\mathrm{Schw}\rho+\frac{1}{2}\frac{\ddot{\rho}^{2}}{\dot{\rho}^{2}}\right)\equiv P.\label{momentum}
\end{equation}
\item Invariance under time translations: $\Delta\rho=-\dot{\rho}$, $F=-L$,
and the conserved quantity is
\begin{equation}
C=-\frac{\dddot{\rho}}{2\dot{\rho}}+\frac{3\ddot{\rho}^{2}}{4\dot{\rho}^{2}}=-\frac{1}{2}{\rm Schw}\rho.
\end{equation}
\item Invariance under space scaling: $\Delta\rho=\rho$, $F=0$, and the
conserved quantity is
\begin{equation}
C=\rho P-\frac{\dddot{\rho}}{2\dot{\rho}}\equiv D.
\end{equation}
\item Invariance under special conformal transformations: $\Delta\rho=\rho^{2}$,
$F=-\dot{\rho}$, and the conserved quantity is
\begin{equation}
C=\rho^{2}P-\frac{\rho\ddot{\rho}}{\dot{\rho}}+\dot{\rho}\equiv K.
\end{equation}
\end{enumerate}
Notice that the
equation of motion (\ref{eqmotion}) can also be written as
\begin{equation}
 \text{EM}=-\frac{\mathrm{d}}{\mathrm{d}t}P=0.
\end{equation}

Now, consider the anisotropic scaling transformation $\Delta\rho=a\rho-b\dot{\rho}t$,
which is a symmetry of the equation of motion, but not of the action since we have $\Delta S=-bS$.  Notice that the quantity $\Lambda=-b$ depends only on the parameter $b$, confirming that the non-Noether part of the anisotropic scaling is associated with this parameter, i.e. $\Delta\rho=-b\dot\rho t$. The corresponding $G$ is
\begin{equation}
G=(a-b)D-\frac{b}{2}t\,{\rm Schw}\rho+b\rho P,
\end{equation}
so, according to the generalized Noether theorem for scaling symmetries (\ref{def-ch}), the conserved quantity $C_S$ is
\begin{equation}
C_{S}=G+bS,
\end{equation}
where $S=\intop_{0}^{t}\mathrm{d}t\,L$ is the on shell action. In order to evaluate the on shell action, observe that the Lagrangian can be written as
\begin{equation}
\frac{\ddot{\rho}^{2}}{4\dot{\rho}^{2}}=\dot{\rho}P-\frac{1}{2}{\rm Schw}\rho.
\end{equation}
Therefore,
\begin{equation}
S=-\intop_{0}^{t}\mathrm{d}t\,\frac{\ddot{\rho}^{2}}{4\dot{\rho}^{2}}=-\intop_{0}^{t}\mathrm{d}t\,\dot{\rho}P+\frac{1}{2}\intop_{0}^{t}\mathrm{d}t\,{\rm Schw}\rho.
\end{equation}
But, since we are evaluating everything on shell, we know that $P=const.$ and ${\rm Schw}\rho=\lambda=const.$, so we have
\begin{equation}
S=-P(\rho-\rho_{0})+\frac{\lambda}{2}t,
\end{equation}
with $\rho_{0}=\rho(t=0)$. Taking  $\rho_{0}=0$ we find the constant of motion $C_{S}$ 
\begin{equation}
C_{S}=(a-b)D-\frac{b}{2}t\,{\rm Schw}\rho+b\rho P+b\left(-\rho P+\frac{\lambda}{2}t\right),
\end{equation}
which simplifies to
\begin{equation}
C_{S}=(a-b)D.
\end{equation}

This example has an interesting property: the scaling symmetry consists of two symmetries, a Noether space scaling given by $\Delta\rho=a\rho$,
whose generator is
\be
D=\frac12\left(2\rho P-\frac{\ddot\rho}{\dot\rho}\right),
\ee
and a scaling in time given by $\Delta\rho=-b\dot\rho t$ that is NOT a symmetry of the action, but is rather a symmetry of the equation of motion. According to our discussion in section 2, its generator is

\begin{equation}
G=-b D-\frac{b}{2}t\,{\rm Schw}\rho+b\rho P,
\end{equation}
(we are restricting ourselves to the case $a=0$). This generator is NOT a constant of motion. Nevertheless, it generates the scaling symmetry $\Delta\rho=-b\dot\rho t$ through the inverse of Noether theorem
\be
\Delta\rho=-2\dot\rho^2\frac{\partial G}{\partial\dddot\rho}=-b\dot\rho t.
\ee
The generator $G$ has also the interesting property
\be
{\frac{ \bar d}{dt}} G=b\,L,
\ee
confirming eq. (\ref{G-eq}).
It is worth noting that the application of the inverse of Noether theorem to $C_S\sim D$ produces the Noether scaling $\Delta\rho\sim\rho$. This is a neat and clear example of our results discussed in section 2.

\subsection{Schwarzian mechanics in the extended formalism}

Schwarzian mechanics in the extended space is a beautiful example of the power of our approach in the context of scaling maps (that may be Noether or non-Noehther symmetries) and conserved currents associated with the symmetries through the extended Noether theorem proposed in our previous section.

The starting point is the Lagrangian
$$L=-\frac14 (\ddot\rho/\dot\rho)^2,$$
and the associated extended Lagrangian
$$\gamma=\frac{\ddot\rho^2 \dot\eta}{2 \dot\rho^3}-\frac{\ddot\rho\ddot \eta}{2
  \dot \rho^2}.$$
The equations of motion are\footnote{The application of the extended space method to this example must be taken with care. This is because the equations of motion in configuration space and the Jacobi equations are not independent in the sense that the equation of motion for $\eta$ depends on the ``accelerations'' of the configuration space $ \rho ^{(4)}(t)  $, so the definition of ``force'' is ambiguous off shell. Nevertheless, our method is able to construct a consistent conserved quantity and symmetries. This fact suggests that the construction of Dirac's method of constrained dynamics in the extended space could be an interesting research  (work in progress).}
$$
 \text{EM}^\eta(\gamma)=-\frac{3 \ddot\rho^3+\ddddot\rho\dot\rho^2-4 \dddot\rho \dot\rho\ddot \rho}{2 \dot\rho^4}=0,
$$
and
$$
 \text{EM}^\rho(\gamma)=\frac{1}{2 \dot\rho^5}\Big(-\ddddot\eta\dot\rho^3+12 \ddot\rho^3 \dot\eta-3 \dot\rho\ddot \rho
 \left(3\ddot \rho \ddot\eta+4 \dddot\rho \dot\eta\right) +
   $$
   $$2 \dot\rho^2 \left(2 \dddot\eta \ddot\rho+2 \dddot\rho\ddot \eta+\ddddot\rho \dot\eta\right)\Big)=0.
   $$
      
   
   In terms of the symmetry $\Delta_s\rho,\Delta_s\eta$ in the extended space we have an associated conserved quantity given by 
\bea
C^E_s&=&
b \Bigg(-\frac{\ddot\rho \dot\eta}{2 \dot\rho^2}+
\eta P+ t \gamma\Bigg)+
   \left(-\frac{ \ddot\eta}{2 \dot\rho^2}+\frac{
    \ddot\rho \dot\eta}{\dot\rho^3}\right)\Delta \dot\rho+\nonumber\\
  &&\frac{1}{\dot\rho^2} \left( \frac{\dddot\eta}{2}-\frac{\ddot\rho \ddot\eta}{\dot\rho}- 
  \dot\eta\,{\rm Schw}\rho
   \right)\Delta\rho-\frac{ \ddot\rho}{2 \dot\rho^2}\Delta\dot\eta+P\Delta\eta.
\eea
Substituting the scaling symmetry given by
\be
\Delta_s\rho=a\rho-b\dot\rho t, \quad\Delta_s\eta=a\eta-b\dot\eta t,
\ee
we can write the constant of motion in the extended space as
$$
C^E_s=\frac{1}{2 \dot\rho^4}\Bigg(\dot\rho^3 \left((b-a) \ddot\eta-b t \dddot\eta\right)-
\dot\rho \left(\ddot\rho^2 \left((a+b) \eta+3 b t \dot\eta\right)+2 a \rho
   \left(\ddot\rho \ddot\eta+\dddot\rho \dot\eta\right)\right)+
   $$
   $$
   \dot\rho^2 \left(\dot\eta \left((a-2 b) \ddot\rho+b t \dddot\rho\right)+(a+b)
   \dddot\rho \eta+a \rho \dddot\eta+3 b t \ddot\rho \ddot\eta\right)+
   3 a \rho \ddot\rho^2 \dot\eta\Bigg).
$$

To verify that this quantity is in fact a constant of motion we observe that
$$
 \frac{d C^E_s}{dt}=-\text{EM} \Big(a \eta-b t  \dot\eta
  +b \eta\Big)- \text{EMJ} \Big(a \rho-b t  \dot \rho\Big),
 $$
where  $\text{EM}$ is the equation of motion in configuration space ($ \text{EM}^\eta(\gamma)$) and EMJ is the Jacobi equation ($ \text{EM}^\rho(\gamma)$). This central result has interesting consequences for the definition of scaling symmetries through the so called inverse of Noether theorem, as was stated in the main text (see section 3).
 
 \subsection{Conserved quantities in configuration space from $C^E_s$}

Now, the projection $\eta\to 1$ (translation) gives
$$
C^E_s\big|_{\eta=1}=\frac{(a+b) \dddot\rho \dot\rho^2-(a+b) \dot\rho\ddot\rho^2}{2\dot\rho^4}.
$$
Recalling that the momentum is defined by
\be
P=\frac12\left(\frac{\dddot\rho}{\dot\rho^2}-\frac{\ddot\rho^2}{\dot\rho^3}\right),
\ee
we find that
$$C^E_s\big|_{\eta=1}=(a+b) P.
$$

Now, projecting on $\eta\to a\rho$ (space scaling), we find that

$$
C^E_s\big|_{\eta=a\rho}=\frac{a b \left(\rho\,\dot \rho\,{\dddot \rho}-\ddot\rho \left(\rho\ddot \rho+\dot\rho^2\right)\right)}{2\dot \rho^3}.
$$
Observing the the dilatation generator is
\be
D=\rho P-\frac{\ddot\rho}{2\dot \rho},
\ee
we find
$$C^E_s\big|_{\eta=a\rho}=ab D.$$

Moreover, projecting on $\eta\to -bt\dot\rho$ (time scaling), we obtain
$$C^E_s\big|_{\eta=-bt\dot\rho}=-ab D.$$
This symmetry is non-Noetherian, nevertheless our formalism is able to handle this case without problems. The complete scaling symmetry $\Delta_s\rho=a\rho-bt\dot\rho$ can not be used to obtain new information, according to the intuition that the deformation of a constant of motion along its own Noetherian symmetry is in general zero.

 If we project on the special conformal symmetry $\eta\to \rho^2$, we obtain 
$$C^E_s\big|_{\epsilon=\rho^2}= -(a-b) K,$$
where $K$ is
$$K=\dot\rho-\frac{\rho \ddot\rho}{\dot\rho}+\frac{\rho^2 \left(\dddot\rho
  \dot \rho-\ddot\rho\right)}{2 \dot\rho^3},$$
  which corresponds to the generator of conformal transformations.
   
    Finally, invariance under time translations  $\eta\to -\dot\rho$ gives
  $$C^E_s\big|_{\eta=-\dot\rho}=-b\,\,\text{Schw}(\rho),$$
   which is consistent with the our previous result in configuration space.\\

    \subsection{On the inverse of Noether theorem in Schwarzian mechanics}

The aim of this subsection is to remark that in some particular cases the application
of the inverse of Noether theorem is not straightforward.  This is the case of our previous example, the Schwarzian mechanics.
 First, observe that $\dot C^E_s$ can be written as
  $$
 \dot C^E_s=-\text{EM} (a \eta -b t  \eta
 +b \eta)- \text{EMJ} (a \rho-b t \dot \rho).
 $$
In the extended space, the equations of motion in configuration space are related (this in turn implies that some constraints may appear in the corresponding Hamiltonian analysis (see previous footnote). The explicit relation is
   $$
     \frac{2\dot\eta}{\dot \rho} \text{EM} + \text{EMJ}-\text{EMJ}\Big|_{\text{EM}=0}=0,
   $$
   where $ \text{EMJ}\Big|_{ \text{EM}=0}$ is the Jacobi equation on shell, i.e. the Jacobi equation evaluated over the equation of motion in configuration space. Thus, we have the relation
   $$
 \dot C^E_s=-\text{EM} \left(\Delta_s\eta+b \eta-  \frac{2\dot\eta}{\dot \rho}\Delta_s\rho\right)- \text{EMJ}\Big|_{\text{EM}=0} \Delta_s\rho.
 $$

 The equations of motion $\text{EM} $ and $\text{EMJ}\Big|_{\text{EM}=0} $ are now independent and we can apply the inverse of Noether theorem to obtain
   \begin{equation}
\frac{d C^E_s}{dt}=\frac{\partial C^E_s}{\partial \dddot\eta }\text{EMJ}\Big|_{\text{EM}=0}
+\frac{\partial C^E_s}{\partial \dddot\rho }\text{EM}.
\end{equation}
Consequently, we have
$$\frac{\partial C^E_s}{\partial \dddot\rho}=\Delta_s\eta+b \eta-  \frac{2\dot\eta}{ \dot\rho}\Delta_s\rho,$$
$$\frac{\partial C^E_s}{\partial \dddot\eta }=\Delta_s\rho,$$
which coincides with the inverse of Noether theorem as reported in section 3, up to the the term $\frac{2\eta '(t)}{ \rho '(t)}\Delta_s\rho$, whose origin was now explained.

\subsection{KdV equation}

Consider the $1+1$ field theory with action
\begin{equation}
S=\int\mathrm{d}t\mathrm{d}\sigma\,\left(-\frac{1}{2}\dot{\phi}\phi^{\prime}+\frac{1}{3}\phi^{\prime3}-\frac{1}{6}\phi^{\prime\prime2}\right).\label{KdVaction}
\end{equation}
Here, the coordinates are $x^{\mu}=(x^{0},x^{1})=(t,\sigma)$, and
the dot and prime denote differentiation with respect to $t$ and
$\sigma$, respectively. The corresponding equation of motion is
\begin{equation}
 \text{EM}=-\frac{1}{3}\phi^{\prime\prime\prime\prime}-2\phi^{\prime}\phi^{\prime\prime}+\dot{\phi}^{\prime}=0,\label{KdVeq}
\end{equation}

Before listing the symmetries of the action (\ref{KdVaction}), we
first recall that for a field theory with Lagrangian ${\cal L}={\cal L}(\phi,\partial_{\mu}\phi,\partial_{\mu}\partial_{\nu}\phi)$,
which under the variation $\Delta\phi$ remains invariant up to a
divergence, i.e., $\Delta{\cal L}=\partial_{\mu}F^{\mu}$, the Noether
theorem reads
\begin{equation}
 \text{EM}\,\Delta\phi+\partial_{\mu}(G^{\mu}-F^{\mu})=0,\label{Noether}
\end{equation}
where $ \text{EM}$ is the equation of motion
\begin{equation}
 \text{EM}=\partial_{\mu}\partial_{\nu}\frac{\partial{\cal L}}{\partial(\partial_{\mu}\partial_{\nu}\phi)}-\partial_{\mu}\frac{\partial{\cal L}}{\partial(\partial_{\mu}\phi)}+\frac{\partial{\cal L}}{\partial\phi},
\end{equation}
and $G^{\mu}$ is
\begin{equation}
G^{\mu}=\left(\frac{\partial{\cal L}}{\partial(\partial_{\mu}\phi)}-\partial_{\nu}\frac{\partial{\cal L}}{\partial(\partial_{\mu}\partial_{\nu}\phi)}\right)\Delta\phi+\frac{\partial{\cal L}}{\partial(\partial_{\mu}\partial_{\nu}\phi)}\partial_{\nu}(\Delta\phi).\label{genfield}
\end{equation}
When the equation of motion holds, there exists a conserved
current $C^{\mu}=G^{\mu}-F^{\mu}$.

In our case, the Lagrangian ${\cal L}=-\frac{1}{2}\dot{\phi}\phi^{\prime}+\frac{1}{3}\phi^{\prime3}-\frac{1}{6}\phi^{\prime\prime2}$,
Eq. (\ref{genfield}) takes the form
\begin{align}
G^{0} & =-\frac{1}{2}\phi^{\prime}\Delta\phi,\nonumber \\
G^{1} & =\left(-\frac{1}{2}\dot{\phi}+\phi^{\prime2}+\frac{1}{3}\phi^{\prime\prime\prime}\right)\Delta\phi-\frac{1}{3}\phi^{\prime\prime}(\Delta\phi)^{\prime},\label{genKdV}
\end{align}
which now allows us to find the symmetries and the corresponding
conserved currents of the KdV action (\ref{KdVaction}).\\

1. Invariance under field displacements: $\Delta\phi=1$, $F^{\mu}=0$.
The conserved current is
\begin{align}
C^{0} & =-\frac{1}{2}\phi^{\prime},\nonumber \\
C^{1} & =-\frac{1}{2}\dot{\phi}+\phi^{\prime2}+\frac{1}{3}\phi^{\prime\prime\prime}.\label{fieldispKdV}
\end{align}

Notice that Eq. (\ref{Noether}) takes the form
\begin{equation}
\partial_{\mu}C^{\mu}=- \text{EM},
\end{equation}

which means, as is well-known, that the KdV equation is a conservation
law of its own action.\\

2. Invariance under space translations: $\Delta\phi=-\phi^{\prime}$,
$F^{0}=0$, $F^{1}=-{\cal L}$. The conserved current is
\begin{align}
C^{0} & =\frac{1}{2}\phi^{\prime2},\nonumber \\
C^{1} & =-\frac{2}{3}\phi^{\prime3}+\frac{1}{6}\phi^{\prime\prime2}-\frac{1}{3}\phi^{\prime}\phi^{\prime\prime\prime}.
\end{align}

3. Invariance under time translations: $\Delta\phi=-\dot{\phi}$,
$F^{0}=-{\cal L}$, $F^{1}=0$. The conserved current is
\begin{align}
C^{0} & =\frac{1}{3}\phi^{\prime3}-\frac{1}{6}\phi^{\prime\prime2},\nonumber \\
C^{1} & =\frac{1}{2}\dot{\phi}^{2}-\dot{\phi}\phi^{\prime2}-\frac{1}{3}\dot{\phi}\phi^{\prime\prime\prime}+\frac{1}{3}\dot{\phi}^{\prime}\phi^{\prime\prime}.
\end{align}

4. Invariance under Galilean boosts: $\Delta\phi=\frac{\sigma}{2}+t\phi^{\prime}$,
$F^{0}=-\frac{1}{4}\phi$ and $F^{1}=t{\cal L}$. The conserved current
is
\begin{align}
C^{0} & =\frac{1}{4}\phi-\frac{1}{4}\sigma\phi^{\prime}-\frac{1}{2}t\phi^{\prime2},\nonumber \\
C^{1} & =-\frac{1}{4}\sigma\dot{\phi}+\frac{2}{3}t\phi^{\prime3}+\frac{1}{2}\sigma\phi^{\prime2}+\frac{1}{6}\sigma\phi^{\prime\prime\prime}+\frac{1}{3}t\phi^{\prime}\phi^{\prime\prime\prime}-\frac{1}{6}\phi^{\prime\prime}-\frac{1}{6}t\phi^{\prime\prime2}.
\end{align}
\\

Furthermore, there is an important symmetry of the equation of motion
(\ref{KdVeq}), which is not a symmetry of the action (\ref{KdVaction}):
the scaling transformation
\begin{equation}
\phi\rightarrow\lambda^{-1}\phi,\,\,\,\,\,\sigma\rightarrow\lambda\sigma,\,\,\,\,\,t\rightarrow\lambda^{3}t,\label{scalingKdV}
\end{equation}
whose infinitesimal version is $\Delta_{s}\phi=-\phi-3\dot{\phi}t-\phi^{\prime}\sigma$,
and applied to the Lagrangian yields $\Delta_{s}{\cal L}=-2{\cal L}-\partial_{t}(3t{\cal L})-\partial_{\sigma}(\sigma{\cal L})$.
Even though this transformation does not correspond to a Noether symmetry,
we can find the function $C^{\mu}=G^{\mu}-F^{\mu}$ which has the
components
\begin{align}
C^{0}= & \frac{1}{2}\phi\phi^{\prime}+\frac{1}{2}\phi^{\prime2}\sigma-\frac{1}{2}\phi^{\prime\prime2}t+\phi^{\prime3}t,\nonumber \\
C^{1}= & \frac{1}{2}\phi\dot{\phi}-\phi\phi^{\prime2}-\frac{1}{3}\phi\phi^{\prime\prime\prime}+\frac{3}{2}\dot{\phi}^{2}t-3\phi^{\prime2}\dot{\phi}t-\phi^{\prime\prime\prime}\dot{\phi}t\nonumber \\
 & -\frac{1}{3}\phi^{\prime}\phi^{\prime\prime\prime}\sigma-\frac{2}{3}\phi^{\prime3}\sigma+\frac{1}{6}\phi^{\prime\prime2}\sigma+\frac{2}{3}\phi^{\prime}\phi^{\prime\prime}+\dot{\phi}^{\prime}\phi^{\prime\prime}t.\label{G1}
\end{align}
Thus, the conserved current obtained through the generalized Noether
theorem for scaling symmetries \cite{Horvathy} formally reads
\begin{equation}
C_{S}^{\mu}=C^{\mu}+2S^{\mu},
\end{equation}
where $S^{\mu}$ is defined as
\begin{equation}
S^{0}=\frac{1}{2}\intop_{0}^{t}\mathrm{d}t^{\prime}\,{\cal L},\,\,\,\,\,S^{1}=\frac{1}{2}\intop_{0}^{\sigma}\mathrm{d}\sigma^{\prime}\,{\cal L},
\end{equation}
which have the property $\partial_{\mu}S^{\mu}={\cal L}$. Notice
that according to \cite{Horvathy}, after the integration of the
Lagrangian, the functions $S^{\mu}$ must be re expressed in terms
of the field $\phi$ and its derivatives.

The inverse of Noether theorem for the KdV action (\ref{KdVaction}) is
\begin{equation}
\Delta\phi=3\frac{\partial G^{1}}{\partial\phi^{\prime\prime\prime}}.
\end{equation}
The symmetry generated by $G^1$, Eq. (\ref{genKdV})
gives
\begin{equation}
\Delta_{s}\phi=-\phi-\sigma\phi^{\prime}-3t\dot{\phi},
\end{equation}
which is precisely the scaling symmetry (\ref{scalingKdV}). However, it is important to
have in mind that in this case the generator (\ref{G1}) is not a conserved quantity.

\subsection{KdV equation in the extended formalism}

The vertical extension of a field theory with Lagrangian ${\cal L}={\cal L}(\phi,\partial_{\mu}\phi,\partial_{\mu}\partial_{\nu}\phi)$
is
\begin{equation}
\gamma=\frac{\partial{\cal L}}{\partial\phi}\eta+\frac{\partial{\cal L}}{\partial(\partial_{\mu}\phi)}\partial_{\mu}\eta+\frac{\partial{\cal L}}{\partial(\partial_{\mu}\partial_{\nu}\phi)}\partial_{\mu}\partial_{\nu}\eta.
\end{equation}
If we consider a scaling transformation $\Delta_{s}\phi$ such that
$\Delta_{s}{\cal L}=\Lambda{\cal L}+\partial_{\mu}F^{\mu}$, then,
following the same steps of Section 3, we find the associated conserved
current in the extended space
\begin{align}
C_{s}^{\mu}= & \left(\frac{\partial\gamma}{\partial(\partial_{\mu}\phi)}-\partial_{\nu}\frac{\partial\gamma}{\partial(\partial_{\mu}\partial_{\nu}\phi)}\right)\Delta_{s}\phi+\frac{\partial\gamma}{\partial(\partial_{\mu}\partial_{\nu}\phi)}\partial_{\nu}(\Delta_{s}\phi)\nonumber \\
 & +\left(\frac{\partial\gamma}{\partial(\partial_{\mu}\eta)}-\partial_{\nu}\frac{\partial\gamma}{\partial(\partial_{\mu}\partial_{\nu}\eta)}\right)\Delta_{s}\eta+\frac{\partial\gamma}{\partial(\partial_{\mu}\partial_{\nu}\eta)}\partial_{\nu}(\Delta_{s}\eta)\nonumber \\
 & -\Lambda\left(\frac{\partial{\cal L}}{\partial(\partial_{\mu}\phi)}\eta-\partial_{\nu}\frac{\partial{\cal L}}{\partial(\partial_{\mu}\partial_{\nu}\phi)}\eta+\frac{\partial{\cal L}}{\partial(\partial_{\mu}\partial_{\nu}\phi)}\partial_{\nu}\eta\right)+\gamma\delta x^{\mu}.
\end{align}

In the case of the KdV action, the corresponding extended Lagrangian
is
\begin{equation}
\gamma=-\frac{1}{2}\phi^{\prime}\dot{\eta}+\left(\phi^{\prime2}-\frac{1}{2}\dot{\phi}\right)\eta^{\prime}-\frac{1}{3}\phi^{\prime\prime}\eta^{\prime\prime}.
\end{equation}
Performing the variation of $\gamma$ with respect to $\eta$, we
find the KdV equation
\begin{equation}
 \text{EM}^{\eta}(\gamma)=-\frac{1}{3}\phi^{\prime\prime\prime\prime}-2\phi^{\prime}\phi^{\prime\prime}+\dot{\phi}^{\prime},
\end{equation}
whereas varying with respect to $\phi$, we find the Jacobi equation \cite{Olver}
\begin{equation}
 \text{EM}^{\phi}(\gamma)=-\frac{1}{3}\eta^{\prime\prime\prime\prime}-2\phi^{\prime}\eta^{\prime\prime}-2\phi^{\prime\prime}\eta^{\prime}+\dot{\eta}^{\prime}.\label{JacobiKdV}
\end{equation}
As we mentioned earlier, the scaling transformation is $\Delta_{s}\phi=-\phi-3\dot{\phi}t-\phi^{\prime}\sigma$,
scales the action (\ref{KdVaction}) with a factor $\Lambda=-2$.
Using the extension of the symmetry $\Delta_{s}\eta=-\eta-3\dot{\eta}t-\eta^{\prime}\sigma$,
the constant of motion $C_{s}^{\mu}$ in the extended formalism is
\begin{align}
C_{s}^{0}= & \frac{1}{2}\eta^{\prime}\phi-\frac{1}{2}\phi^{\prime}\eta+\eta^{\prime}\phi^{\prime}\sigma+3\phi^{\prime2}\eta^{\prime}t-\phi^{\prime\prime}\eta^{\prime\prime}t,\nonumber \\
C_{s}^{1}= & \frac{1}{2}\dot{\eta}\phi-\frac{1}{2}\dot{\phi}\eta+3\dot{\phi}\dot{\eta}t-2\phi\phi^{\prime}\eta^{\prime}-2\phi^{\prime2}\eta^{\prime}\sigma-6\dot{\phi}\phi^{\prime}\eta^{\prime}t\nonumber \\
 & +\frac{2}{3}\phi^{\prime}\eta^{\prime\prime}+\eta^{\prime\prime}\dot{\phi}^{\prime}t-\frac{1}{3}\eta^{\prime\prime\prime}(\phi+3\dot{\phi}t+\phi^{\prime}\sigma)+\phi^{\prime2}\eta-3\phi^{\prime2}\dot{\eta}t\nonumber \\
 & +\frac{1}{3}\phi^{\prime\prime}\eta^{\prime\prime}\sigma+\phi^{\prime\prime}\dot{\eta}^{\prime}t+\frac{1}{3}\phi^{\prime\prime\prime}(\eta-3\dot{\eta}t-\eta^{\prime}\sigma),\label{extconstant}
\end{align}
and its off-shell divergence is
\begin{equation}
\partial_{\mu}C_{s}^{\mu}=-\text{EMJ}(-\phi-3\dot{\phi}t-\phi^{\prime}\sigma)- \text{EM}(\eta-3\dot{\eta}t-\eta^{\prime}\sigma),
\end{equation}
where $ \text{EM}$ is the KdV equation (\ref{KdVeq}) and $\text{EMJ}$ is the Jacobi
equation (\ref{JacobiKdV}).

Finally, the inverse of Noether theorem for the scaling symmetry of
the KdV equation is quite simple and reads
\begin{equation}
\Delta_{s}\phi=3\frac{\partial C^{1}}{\partial\eta^{\prime\prime\prime}},\,\,\,\,\,\Delta_{s}\eta+2\eta=3\frac{\partial C^{1}}{\partial\phi^{\prime\prime\prime}},
\end{equation}
which has the expected form because the scaling factor of the action
is $\Lambda=-2$. Notice that in contrast to the Schwarzian mechanics, there are no extra terms here since the Jacobi equation (\ref{JacobiKdV}) does not depend on the highest derivative of the field, $\phi^{\prime\prime\prime\prime}$.

\subsection{Conserved quantities in the configuration space from $C_{s}^{\mu}$}

Let us now project the constant of motion $C_{s}^{\mu}$ to configuration
space by considering the following symmetries of the action:\\

1. Invariance under field displacements: $\eta=1$. The resulting
projection is
\begin{align}
C_{s}^{0}\bigg|_{\eta=1} & =-\frac{1}{2}\phi^{\prime},\nonumber \\
C_{s}^{1}\bigg|_{\eta=1} & =-\frac{1}{2}\dot{\phi}+\phi^{\prime2}+\frac{1}{3}\phi^{\prime\prime\prime},
\end{align}

which coincides with Eq. (\ref{fieldispKdV}).

2. Invariance under space translations: $\eta=-\phi^{\prime}$. The
resulting projection is
\begin{align}
C_{s}^{0}\bigg|_{\eta=-\phi^{\prime}}= & \frac{1}{2}\phi^{\prime2}-\frac{1}{2}\phi\phi^{\prime\prime}-\sigma\phi^{\prime}\phi^{\prime\prime}-3t\phi^{\prime2}\phi^{\prime\prime}+t\phi^{\prime\prime}\phi^{\prime\prime\prime},\nonumber \\
C_{s}^{1}\bigg|_{\eta=-\phi^{\prime}}= & -\phi^{\prime3}+2\phi\phi^{\prime}\phi^{\prime\prime}+2\sigma\phi^{\prime2}\phi^{\prime\prime}-\phi^{\prime}\phi^{\prime\prime\prime}+\frac{1}{3}\phi^{\prime\prime\prime\prime}(\phi+3\dot{\phi}t+\phi^{\prime}\sigma)\nonumber \\
 & +\frac{1}{2}\phi\dot{\phi}+6t\phi^{\prime}\phi^{\prime\prime}\dot{\phi}-\frac{1}{2}\phi\dot{\phi}^{\prime}+3t\phi^{\prime2}\dot{\phi}^{\prime}-3t\phi^{\prime}\dot{\phi}^{\prime}-t\phi^{\prime\prime}\dot{\phi}^{\prime\prime}.\label{cextspace}
\end{align}

After some manipulation, we can extract the associated Noether current
in the configuration space. We first use the KdV equation to substitute
the fourth spatial derivative in Eq. (\ref{cextspace}), which results
in
\begin{equation}
C_{s}^{1}\bigg|_{\eta=-\phi^{\prime}}=-\phi^{\prime3}-\phi^{\prime}\phi^{\prime\prime\prime}+\frac{1}{2}\phi^{\prime}\dot{\phi}+\frac{1}{2}\phi\dot{\phi}^{\prime}+\sigma\phi^{\prime}\dot{\phi}^{\prime}+3t\phi^{\prime2}\dot{\phi}^{\prime}-t\phi^{\prime\prime}\dot{\phi}^{\prime\prime}.\label{cextspaceonshell}
\end{equation}
Now, interestingly enough, we notice that the first component in (\ref{cextspace})
can be rewritten as
\begin{equation}
C_{s}^{0}\bigg|_{\eta=-\phi^{\prime}}=3\left(\frac{1}{2}\phi^{\prime2}\right)+\partial_{\sigma}A,\label{c0space}
\end{equation}
where we have defined $A\equiv-\frac{1}{2}\phi\phi^{\prime}-\frac{1}{2}\phi^{\prime2}\sigma-\phi^{\prime3}t+\frac{1}{2}\phi^{\prime\prime2}t$.
Moreover, from (\ref{cextspaceonshell}), we see that
\begin{equation}
C_{s}^{1}\bigg|_{\eta=-\phi^{\prime}}=3\left(-\frac{2}{3}\phi^{\prime3}+\frac{1}{6}\phi^{\prime\prime2}-\frac{1}{3}\phi^{\prime}\phi^{\prime\prime\prime}\right)-\partial_{t}A.\label{c1space}
\end{equation}
From here, it is evident that the terms in parentheses in (\ref{c0space})
and (\ref{c1space}) are $G^{0}$ and $G^{1}$, respectively. Thus,
the on-shell divergence (denoted with a bar) of this projected $C_{s}^{\mu}$
is
\begin{equation}
\bar{\partial}_{\mu}C_{s}^{\mu}\bigg|_{\eta=-\phi^{\prime}}=3\left(\partial_{t}G^{0}+\partial_{\sigma}G^{1}\right)=0,
\end{equation}
where the function $A$ has been canceled due to the symmetry of the
mixed derivatives. In this manner, we recover the original generators,
up to the generator of a trivial symmetry transformation \cite{Olver,Olver-1}.

3. Invariance under time translations: $\eta=-\dot{\phi}$. The resulting
projection is
\begin{align}
C_{s}^{0}\bigg|_{\eta=-\dot{\phi}}= & \frac{1}{2}\dot{\phi}\phi^{\prime}-\frac{1}{2}\phi\dot{\phi}^{\prime}-\sigma\phi^{\prime}\dot{\phi}^{\prime}-3t\phi^{\prime2}\dot{\phi}^{\prime}+t\phi^{\prime\prime}\dot{\phi}^{\prime\prime},\nonumber \\
C_{s}^{1}\bigg|_{\eta=-\dot{\phi}}= & -\phi^{\prime2}\dot{\phi}-\frac{1}{3}\phi^{\prime\prime\prime}\left(\dot{\phi}-3\ddot{\phi}t-\dot{\phi}^{\prime}\sigma\right)+\frac{1}{2}\dot{\phi}^{2}+2\phi\phi^{\prime}\dot{\phi}^{\prime}+2\sigma\phi^{\prime2}\dot{\phi}^{\prime}-\frac{2}{3}\phi^{\prime}\dot{\phi}^{\prime\prime}\nonumber \\
 & +6t\phi^{\prime}\dot{\phi}\dot{\phi}^{\prime}-\frac{1}{3}\sigma\phi^{\prime\prime}\dot{\phi}^{\prime\prime}-t\dot{\phi}^{\prime}\dot{\phi}^{\prime\prime}+\frac{1}{3}\dot{\phi}^{\prime\prime\prime}\left(\phi+3\dot{\phi}t+\phi^{\prime}\sigma\right)-\frac{1}{2}\phi\ddot{\phi}\nonumber \\
 & +3t\phi^{\prime2}\ddot{\phi}-3t\dot{\phi}\ddot{\phi}-t\phi^{\prime\prime}\ddot{\phi}^{\prime},
\end{align}

and its on-shell divergence vanishes.

4. Invariance under Galilean boosts: $\eta=\frac{\sigma}{2}+t\phi^{\prime}$.
The resulting projection is
\begin{align}
C_{s}^{0}\bigg|_{\eta=\frac{\sigma}{2}+t\phi^{\prime}}= & \frac{1}{4}\phi+\frac{1}{4}\sigma\phi^{\prime}+t\phi^{\prime2}+\frac{1}{2}t\phi\phi^{\prime\prime}+t\sigma\phi^{\prime}\phi^{\prime\prime}+3t^{2}\phi^{\prime2}\phi^{\prime\prime}-t^{2}\phi^{\prime\prime}\phi^{\prime\prime\prime},\nonumber \\
C_{s}^{1}\bigg|_{\eta=\frac{\sigma}{2}+t\phi^{\prime}}= & -\frac{1}{2}\phi\phi^{\prime}-\frac{1}{2}\sigma\phi^{\prime2}-2t\phi^{\prime3}-2t\phi\phi^{\prime}\phi^{\prime\prime}-2t\sigma\phi^{\prime2}\phi^{\prime\prime}+t\phi^{\prime\prime2}-\frac{1}{3}t\phi\phi^{\prime\prime\prime\prime}\nonumber \\
 & -\frac{1}{3}t\sigma\phi^{\prime}\phi^{\prime\prime\prime\prime}-\frac{1}{4}\sigma\dot{\phi}-\frac{1}{2}t\dot{\phi}\phi^{\prime}-6t^{2}\dot{\phi}\phi^{\prime}\phi^{\prime\prime}-t^{2}\dot{\phi}\phi^{\prime\prime\prime\prime}+\frac{1}{2}t\phi\dot{\phi}^{\prime}\nonumber \\
 & -3t^{2}\dot{\phi}^{\prime}\phi^{\prime2}+3t^{2}\dot{\phi}\dot{\phi}^{\prime}+t^{2}\phi^{\prime\prime}\dot{\phi}^{\prime\prime},
\end{align}
and can be corroborated that its on-shell divergence is zero.

We can also construct a conserved current associated with the scaling symmetry but the result is not very illuminating. Nevertheless, a simple observation allows us to construct a different conserved current associated also with scaling symmetry. Notice that the Jacobi equation for the KdV equation (\ref{JacobiKdV}) has a peculiar property: it is the divergence of some current given by
$$G^0:=\eta', \qquad G^1:=-\frac13\eta'''-2\phi'\eta',$$
so the Jacobi equation (\ref{JacobiKdV}) is
$$\partial_t G^0+\partial_\sigma G^1=\text{EMJ}.$$

If we now evaluate this equation on the scaling symmetry $\eta=\phi+3t\dot\phi+\sigma\phi'$, we obtain
$$G^0=2\phi'+3t\dot\phi'+\sigma\phi'',$$
$$ G^1=-4 \phi'^2-\frac43\phi''' -\sigma\dot\phi'-6t\phi'\dot\phi'-t\dot\phi'''.$$
In fact, taking the diverge of this current, we find
$$\text{div} G= \text{EM}+\partial_t(3t  \text{EM})+\partial_\sigma(\sigma  \text{EM}),$$
where $ \text{EM}$ is the KdV equation. Of course, this divergence is zero on-shell $ \text{EM}=0$. 
This current can also be cast in a more suggestive way
$$G^0=2\phi'+3t\dot\phi'+\sigma\phi'' -3t  \text{EM},$$
$$ G^1=-4 \phi'^2-\frac43\phi''' -\sigma\dot\phi'-6t\phi'\dot\phi'-t\dot\phi'''-\sigma  \text{EM},$$
showing the the on shell current is exactly $G^0,G^1$. With this notation,
$$\text{div} G= \text{EM}.$$
This conserved quantity has been constructed some time ago through an elaborated argument in \cite {RS,RK}. KdV is not the only non linear partial differential equation with this property. Many others interesting non linear equations share also this property.

\section*{Acknowledgements} 

RA was partially supported by a PhD. CONACyT fellowship number 744575. DGR is supported with a CONACyT Ph.D. fellowship number 332577. The work of JAG was partially supported CONACyT grant A1-S-22886 and DGAPA-UNAM grant IN107520.

\section{Appendix}

\subsection{A1: Symmetries of the equations of motion}

The infinitesimal symmetries of the equations of motion obey a very interesting relation known as the Jacobi equation, or the second variation of the Lagrangian action, if the system admits a variational formulation. But these symmetries are defined just as symmetries of the equation of motion, independently of whether the equations of motion admit a variational formulation or not. They are defined as infinitesimal transformations that map solutions of the dynamical system into solutions of the dynamical system.
First, we can associate to every infinitesimal symmetry that also moves the time
$$\delta q^i=\bar q^i(\bar t)-q^i(t),\qquad  \delta t=\bar t-t,$$
an infinitesimal symmetry that does not move time, given by
$$\Delta q^i=\delta q^i-\dot q^i \delta t.$$
Now, if we define
$\Delta q^i=\eta^i(q,\dot q,t),$
it will be a symmetry of the equations of motion iff
$$\ddot q^i-F^i(q^j,\dot q^j,t)=0 \quad\Longleftrightarrow\quad
\ddot {\bar q}^i-F^i({\bar q}^j,\dot{\bar q}^j,t)=0,$$
which in turn implies \cite{Hojman}
\begin{equation}\label{sym-eq-m}
\frac{\bar d}{dt}\frac{\bar d}{dt} \eta^i-\frac{\partial F^i}{\partial\dot q^j}\frac{\bar d}{dt}\eta^j- \frac{\partial F^i}{\partial q^j} \eta^j=0.
\end{equation}
So, every symmetry of the equations of motion (including Noether symmetries) are solutions of this equation. In particular, for the scaling transformation $\eta^i=aq^i-b\dot{q}^i t$, the Jacobi equation gives
\begin{equation}
\frac{\bar d}{dt}\frac{\bar d}{dt} (a q^i-b\dot q^i t)-\frac{\partial F^i}{\partial\dot q^j}\frac{\bar d}{dt}(a q^j-b\dot q^j t)- \frac{\partial F^i}{\partial q^j} (a q^j-b\dot q^j t)=0,
\end{equation}
which in turn implies the following condition for the forces of the dynamical system
\begin{equation}\label{force-cond}
-bt\frac{\partial F^i}{\partial t}+(a-2b)F^i-(a-b)  \frac{\partial F^i}{\partial\dot q^j}\dot q^j-\frac{\partial F^i}{\partial q^j} a q^j=0.
\end{equation}
For all the examples that we have considered, these equations are fulfilled. We are unaware if these relations were considered previously in the literature. In the case of AFF conformal mechanics, $F\sim\frac{1}{q^3}$, and $a=1,b=2$, so the equation (\ref{force-cond}) reduces to
$$ \frac{\partial F}{\partial q}  q=-3 F,$$
thus, $F$ must be a homogeneous function of degree $-3$ as expected.

\subsection{A2: The Jacobi equation in terms of the Lagranigian}

The coefficients of the Jacobi equation in terms of the Lagrangian function $L$ are
\be
W_{ij}=\frac{\partial^2 L}{\partial\dot q^i\partial\dot q^j},
\ee
\be
N_{ij}=\frac{d}{dt}\left(\frac{\partial^2 L}{\partial\dot q^i\partial\dot q^j}\right)+\frac{\partial^2 L}{\partial\dot q^i\partial q^j}-\frac{\partial^2 L}{\partial q^i\partial\dot q^j},
\ee
and
\be
M_{ij}=\frac{d}{dt}\left(\frac{\partial^2 L}{\partial\dot q^i\partial q^j}\right)-\frac{\partial^2 L}{\partial q^i\partial q^j}.
\ee

\end{document}